\PassOptionsToPackage{unicode}{hyperref}
\PassOptionsToPackage{hyphens}{url}
\documentclass[
]{article}
\usepackage{amsmath,amssymb}
\usepackage{lmodern}
\usepackage{iftex}
\usepackage{multicol}
\usepackage{wrapfig}
\usepackage{caption} % Required for \captionof
\usepackage{changepage} % Required for adjustwidth
\usepackage[left=2.5cm, right=2.5cm, top=3cm, bottom=3cm]{geometry}
\ifPDFTeX
  \usepackage[T1]{fontenc}
  \usepackage[utf8]{inputenc}
  \usepackage{textcomp} % provide euro and other symbols
\else % if luatex or xetex
  \usepackage{unicode-math}
  \defaultfontfeatures{Scale=MatchLowercase}
  \defaultfontfeatures[\rmfamily]{Ligatures=TeX,Scale=1}
\fi
\IfFileExists{upquote.sty}{\usepackage{upquote}}{}
\IfFileExists{microtype.sty}{% use microtype if available
  \usepackage[]{microtype}
  \UseMicrotypeSet[protrusion]{basicmath} % disable protrusion for tt fonts
}{}
\makeatletter
\@ifundefined{KOMAClassName}{% if non-KOMA class
  \IfFileExists{parskip.sty}{%
    \usepackage{parskip}
  }{% else
    \setlength{\parindent}{0pt}
    \setlength{\parskip}{6pt plus 2pt minus 1pt}}
}{% if KOMA class
  \KOMAoptions{parskip=half}}
\makeatother
\usepackage{xcolor}
\usepackage{longtable,booktabs,array}
\usepackage{calc} % for calculating minipage widths
\usepackage{etoolbox}
\makeatletter
\patchcmd\longtable{\par}{\if@noskipsec\mbox{}\fi\par}{}{}
\makeatother
\IfFileExists{footnotehyper.sty}{\usepackage{footnotehyper}}{\usepackage{footnote}}
\makesavenoteenv{longtable}
\usepackage{graphicx}
\makeatletter
\def\maxwidth{\ifdim\Gin@nat@width>\linewidth\linewidth\else\Gin@nat@width\fi}
\def\maxheight{\ifdim\Gin@nat@height>\textheight\textheight\else\Gin@nat@height\fi}
\makeatother
\setkeys{Gin}{width=\maxwidth,height=\maxheight,keepaspectratio}
\makeatletter
\def\fps@figure{htbp}
\makeatother
\ifLuaTeX
  \usepackage{selnolig}  % disable illegal ligatures
\fi
\IfFileExists{bookmark.sty}{\usepackage{bookmark}}{\usepackage{hyperref}}
\IfFileExists{xurl.sty}{\usepackage{xurl}}{} % add URL line breaks if available
\hypersetup{
  pdftitle={Lidar-monitoring of atmospheric fluctuations for millimeter-wave astronomical observations},
  hidelinks,
  pdfcreator={LaTeX via pandoc}}

\title{\protect\hypertarget{_Toc425433217}{}{} \bf Lidar-monitoring of
atmospheric fluctuations for millimeter-wave astronomical observations}
\author{}
\date{}

\begin{document}
\maketitle

\begin{adjustwidth}{0.15in}{0.15in}

\begin{center}
\textbf{
Jacques Delabrouille\textsuperscript{(a,b)}, Patrick
Chazette\textsuperscript{(c)}, Étienne Burtin\textsuperscript{(d)},
Christophe Chailan\textsuperscript{(e)},
Josquin~Errard\textsuperscript{(e)},
Shamik~Ghosh\textsuperscript{(b,a)},
Manuel~Gonzalez\textsuperscript{(e)}, John Groh\textsuperscript{(b)},
Reijo~Keskitalo\textsuperscript{(b)},
Valérian~Le~Lorec\textsuperscript{(d)},
Sotiris~Loucatos\textsuperscript{(d,e)},
Christophe~Magneville\textsuperscript{(d)},
Jean-Baptiste~Melin\textsuperscript{(d)},
Michel~Piat\textsuperscript{(e)}, Damien~Prêle\textsuperscript{(e)},
Julien~Tang\textsuperscript{(a,b,e)},
Timothée~Tollet\textsuperscript{(e)}, and
Julien~Totems\textsuperscript{(c)}\\
}
\vspace{0.3cm}

{ \small
\textsuperscript{(a)} CNRS-UCB, Centre Pierre Binétruy, IRL 2007,
CPB-IN2P3, Berkeley, CA 94720, USA\\
\textsuperscript{(b)} Lawrence Berkeley National Laboratory, 1 Cyclotron
Road, Berkeley, CA 94720, USA\\
\textsuperscript{(c)} Laboratoire des Sciences du Climat et de
l'Environnement (LSCE), CNRS-CEA-UVSQ, Gif sur Yvette, France \\
\textsuperscript{(d)} Université Paris-Saclay, CEA, IRFU, F-91191
Gif-sur-Yvette, France \\
\textsuperscript{(e)} Université Paris Cité, CNRS, Astroparticule et
Cosmologie, F-75013 Paris, France \\
}
Corresponding author: jacques.delabrouille@apc.in2p3.fr
\end{center}
\end{adjustwidth}

\vspace{0.3cm}

\begin{adjustwidth}{0.5in}{0.5in}
\textbf{Abstract:} Ground-based millimeter-wave astronomical
observations suffer from strong contamination by the emission of
inhomogeneous water vapor in the Earth\textquotesingle s atmosphere.
This emission reduces the observing efficiency and data quality even
from a high-altitude, dry location such as the Atacama plateau in Chile.
We propose to continuously monitor atmospheric water vapor with a
co-pointing dedicated Raman lidar system. We show, with numerical
simulations and proof-of-concept observations, how a synergistic use of
atmospheric lidar data has the potential to substantially improve the
quality of astronomical data sets.
\end{adjustwidth}

\hypertarget{section}{%
\section{}\label{section}}

\begin{multicols}{2}

\hypertarget{introduction}{%
\section{Introduction}\label{introduction}}

Millimeter-wave astronomy is of utmost importance for understanding
astrophysical phenomena across cosmic history. Processes at work in the
first second after the Big-Bang or during the formation of galaxies and
clusters of galaxies that constitute the cosmic web, or in the mixture
of gas, dust and magnetic fields of the interstellar medium of our own
Galaxy, all emit subtle millimeter-wave signatures that are key to
answering some of the most fundamental questions in physics and
astronomy.

Large telescopes and telescope arrays such as the Simons Observatory
(SO) {[}1{]} and the Fred Young Submillimeter Telescope {[}2{]}, located
on the Atacama plateau in Chile, are poised to map large areas of the
sky at frequencies ranging from 30 to 900 GHz. However, as seen in the
data of previous instruments such as POLARBEAR {[}3{]} and the Atacama
Cosmology Telescope {[}4{]}, precipitable water vapor (PWV) fluctuations
along the line of sight are a major source of noise for mapping
degree-scale astronomical features. They typically reduce the large-scale
sensitivity of the observations by an order of magnitude or two
compared to what is expected from instrumental noise alone. This
translates into a significant loss of observing efficiency, which requires
inversely proportional total observing time to achieve a sensitivity
objective. Millimeter observations within the Earth's
atmosphere are illustrated in Figure 1.

Atmospheric lidars are classically used to measure atmospheric water
vapor mixing ratio (WVMR) profiles using laser pulses at 355 nm {[}5{]}.
They can be equipped with rotational Raman, vibrational Raman, and
elastic channels to measure water vapor, temperature, aerosols, and
hydrometeor distributions as a function of time along the line of sight.

\begin{center}
    \nopagebreak % Keeps figure attached to surrounding context if needed
    \includegraphics[width=\linewidth]{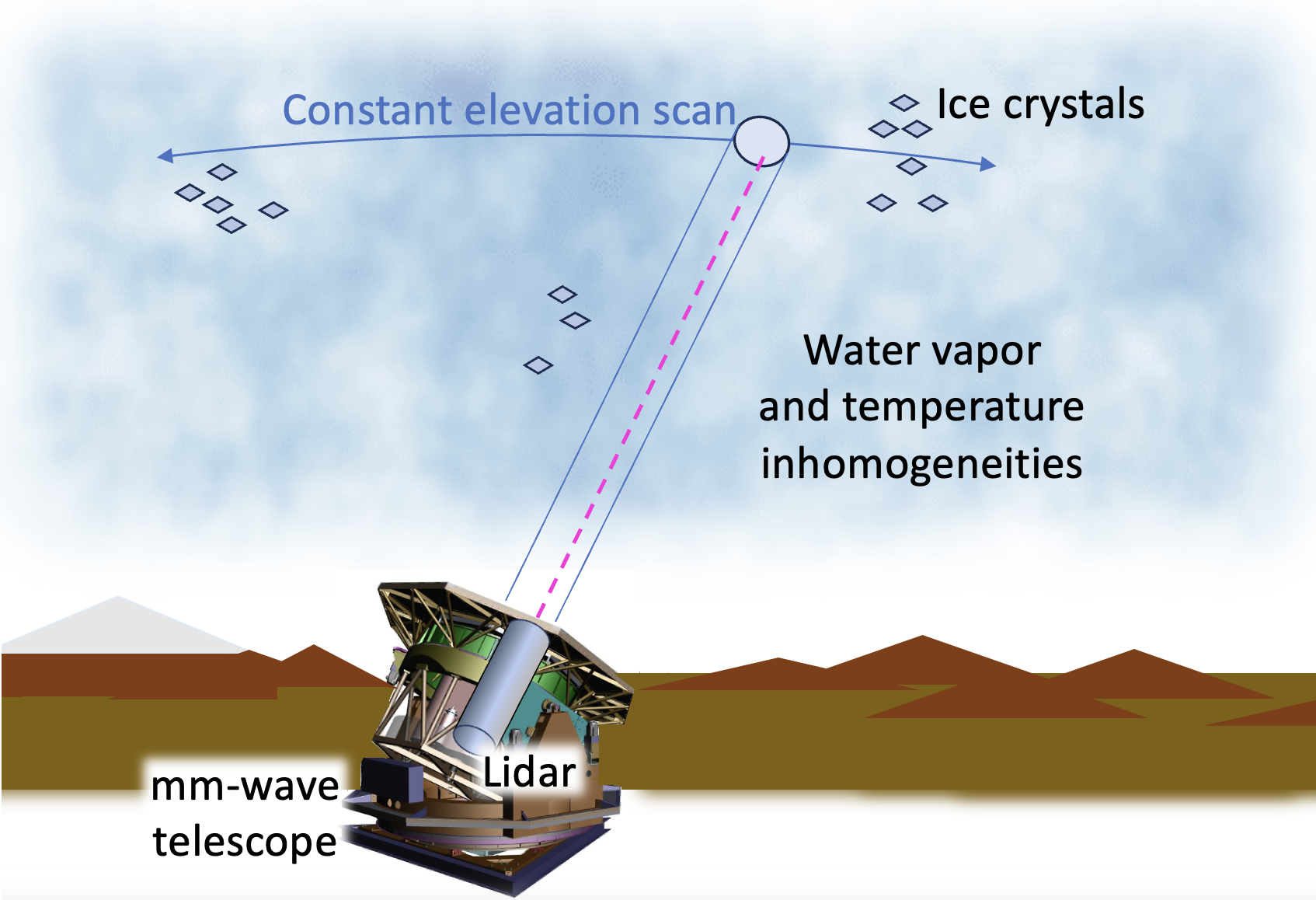}
    \captionof{figure}{Observations within the Earth atmosphere by a mm-wave telescope (design from reference {[}6{]}), and a co-pointing lidar.}
    \label{fig:1}
\end{center}

%\begin{figure}
%\includegraphics[width=\columnwidth]{image1.png}
%\caption{Observations within the Earth atmosphere by a mm-wave telescope
%(design from reference {[}6{]}), and a co-pointing lidar.}
%\label{fig:1}
%\end{figure}

We evaluated the extent to which a lidar system could be used to measure
the atmospheric signal due to PWV fluctuations in millimeter-wave
telescopes and correct the astronomical data streams from this
contamination.

\hypertarget{end-to-end-modeling}{%
\section{End-to-end modeling}\label{end-to-end-modeling}}

This study requires the implementation of end-to-end modeling in order
to design the lidar system. The performance assessment is based on
existing instrumentation that will serve as a basis for the construction
of a future "CosmoLidar" dedicated to millimeter-wave astronomy and
cosmology. Following this dimensioning, which provides access to the
noise level of each channel of the lidar instrument, a simulation of its
measurement within a realistic turbulent atmospheric environment is
carried out to evaluate the suitability and the performance of the lidar
measurement for the correction of millimeter-wave astronomical
observations.

\hypertarget{modeling-assumptions}{%
\subsection{Modeling assumptions}\label{modeling-assumptions}}

We consider an observational setup in which a millimeter telescope scans
the sky in azimuth at a constant elevation of 50°. During the scan, each
detector of the millimeter telescope collects a data stream proportional
to the variations of the total incident radiation coming from its line
of sight in its frequency band. Around 100 and 150 GHz (wavelengths of 3
and 2 mm), in a dry and high-altitude site such as the Atacama plateau,
the dominant signal observed by mm-wave telescopes is roughly
proportional to the variations of integrated precipitable water vapor
(PWV) along the consecutive lines of sight. Measurement of those variations
and subtracting them would significantly improve the sensitivity of
the observations to actual astronomical emissions. Other spurious
signals, due to temperature variations and scattering by ice crystals,
cannot be entirely neglected but are sub-dominant. We do not include
them in the present analysis.

\begin{center}
    \nopagebreak % Keeps figure attached to surrounding context if needed
    \includegraphics[width=\linewidth]{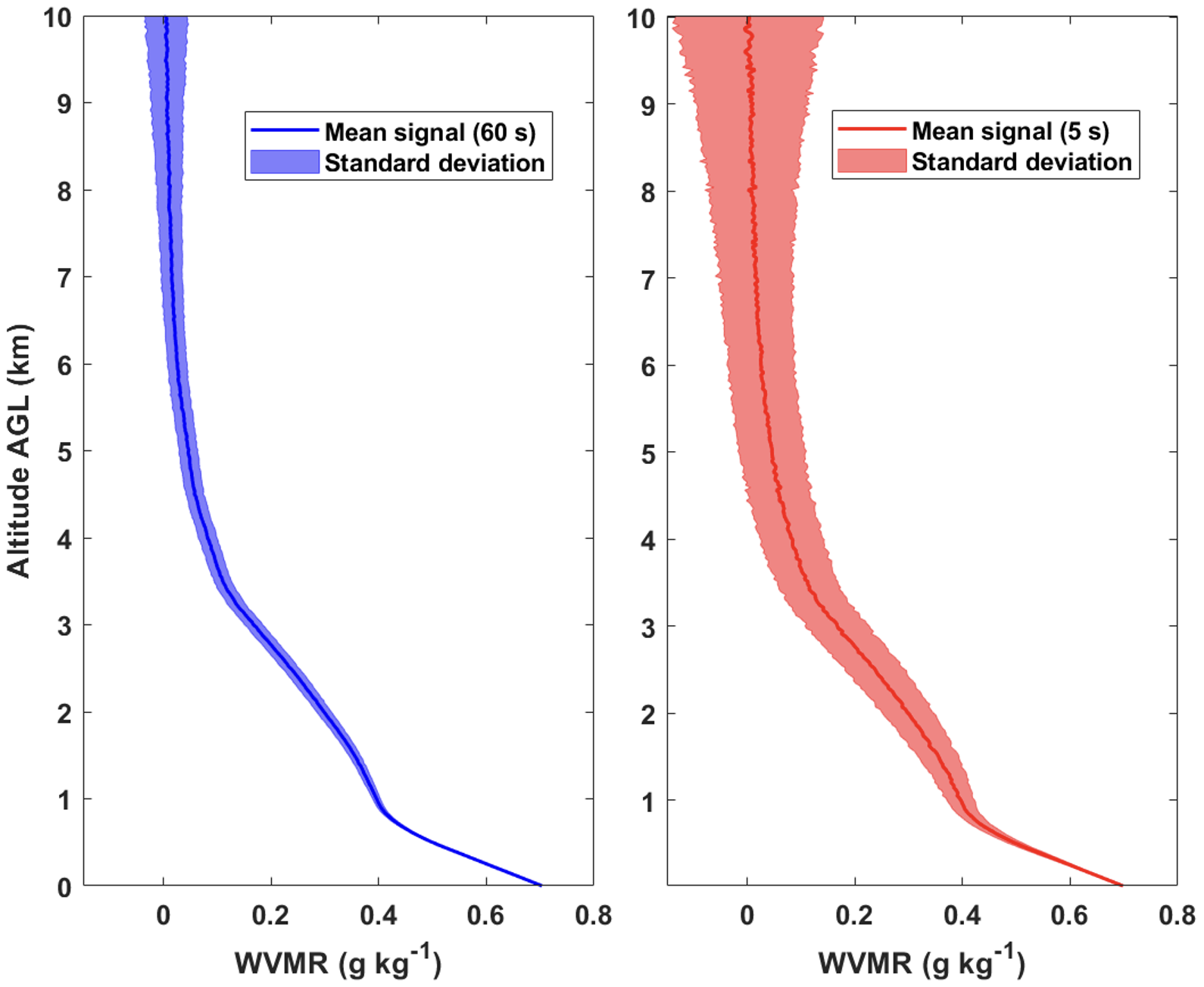}
    \captionof{figure}{Simulation of WVMR profile using ERA5 reanalysis on 16 September 2023 above SO, left with 60 s and right with 5 s integration times, in 30 m altitude bins.}
    \label{fig:2}
\end{center}

%\begin{figure}[tbp]
%\centering\includegraphics[width=\textwidth]{image2.png}
%\caption{Simulation of WVMR profile using ERA5 reanalysis on 16 September 2023 above SO, left with 60 s and right with 5 s integration times, in 30 m altitude bins.}
%\label{fig:2}
%\end{figure}

A noiseless water vapor data stream for a six-minute-long scan spanning
360° in azimuth is simulated by scanning at 50° elevation through an
AtmoCube simulation (see subsection 2.3). We assume that a co-pointing
atmospheric lidar measures the water vapor profiles at each degree of
azimuth along the line of sight. We show in Figure 2 lidar simulations
performed for 60 s integration time and scaled to 5 s.

\hypertarget{initial-lidar-simulation}{%
\subsection{Initial lidar simulation}\label{initial-lidar-simulation}}

The lidar-derived WVMR profiles measured for a set of time samples along the
scan are simulated using the lidar parameters in Table 1. In our
simulation, we use realistic profiles derived from the ERA5 reanalysis
{[}7{]} over the Atacama plateau in Chile on 16 September 2023
(Figure~2). The standard deviations evaluated on these profiles are used
to constrain the noise level in the final end-to-end modeling.

\begin{center}
    \captionof{table}{Lidar simulation parameters}
    \label{tab:multicol_table}
    \begin{tabular}{ll}
        \toprule
        Emitted energy & 60 mJ \\
Pulse repetition frequency & 200 Hz \\  
Reception channels & 386.80 nm
407.65 nm \\
Telescope diameter & 500 mm \\
Field-of-view & 2 mrad \\
Total optical efficiency & 0.24 \\
Dark Count & 15 s\textsuperscript{-1} \\
Vertical resolution & 30 m \\
Temporal resolution & 5 s \\
        \bottomrule
    \end{tabular}
\end{center}

%Table 1. Lidar simulation parameters

%\begin{longtable}[]{@{}
%  >{\raggedright\arraybackslash}p{(\columnwidth - 2\tabcolsep) * \real{0.6088}}
%  >{\raggedright\arraybackslash}p{(\columnwidth - 2\tabcolsep) * \real{0.3912}}@{}}
%\toprule()
%\begin{minipage}[b]{\linewidth}\raggedright
%Emitted energy
%\end{minipage} & \begin{minipage}[b]{\linewidth}\raggedright
%60 mJ
%\end{minipage} \\
%\midrule()
%\endhead
%Pulse repetition frequency & 200 Hz \\
%Reception channels & 386.80 nm
%407.65 nm \\
%Telescope diameter & 500 mm \\
%Field-of-view & 2 mrad \\
%Total optical efficiency & 0.24 \\
%Dark Count & 15 s\textsuperscript{-1} \\
%Vertical resolution & 30 m \\
%Temporal resolution & 5 s \\
%\bottomrule()
%\end{longtable}

\hypertarget{simulations-with-atmocube}{%
\subsection{Simulations with AtmoCube}\label{simulations-with-atmocube}}

On small physical scales, the amount of water vapor fluctuates in the
three-dimensional troposphere as turbulent vertices transport air with
different concentrations of water vapor between different atmospheric
layers. These atmospheric perturbations are the main source of the
low-frequency noise excess for millimeter astronomy. This kind of
turbulent process generates a cascade of perturbations with a 3D
Kolmogorov spectrum \(P(k)\) given by

\[P(k) \propto \ \left( \frac{\ k_{\min}^{2} + k^{2}}{k_{\min}^{2}} \right)^{- \frac{11}{6}},\]

where \(k_{\min}\) is an injection scale. We developed a simulation
tool, AtmoCube, to generate cubes \(\delta(x,y,z)\) of random
fluctuations of 1024\textsuperscript{3} ten-meter cells according to
this \(P(k)\), scaled as a function of altitude to get

\[\delta WVMR(x,y,z)\  = \ \overline{\ \delta WVMR}(z) \times \delta(x,y,z),\]

where \(\overline{\ \delta WVMR}(z)\) is a mean profile for the
amplitude of the WVMR fluctuations, assumed to be proportional to the
WVMR itself. For the specific simulation in the present study, we use a
profile inferred from the ERA5 reanalysis data on 16 September 2023 at
0h UTC, above the Simons Observatory located at an altitude of 5190~m on
the Atacama plateau in Chile, at a longitude of 67.79°W and a latitude
of 22.96°S. The amplitude of the fluctuations is set to match the actual
observations of the integrated PWV fluctuations at that particular time.
A down-sampled cube is shown in Figure~3.

\begin{center}
    \nopagebreak % Keeps figure attached to surrounding context if needed
    \includegraphics[width=\linewidth]{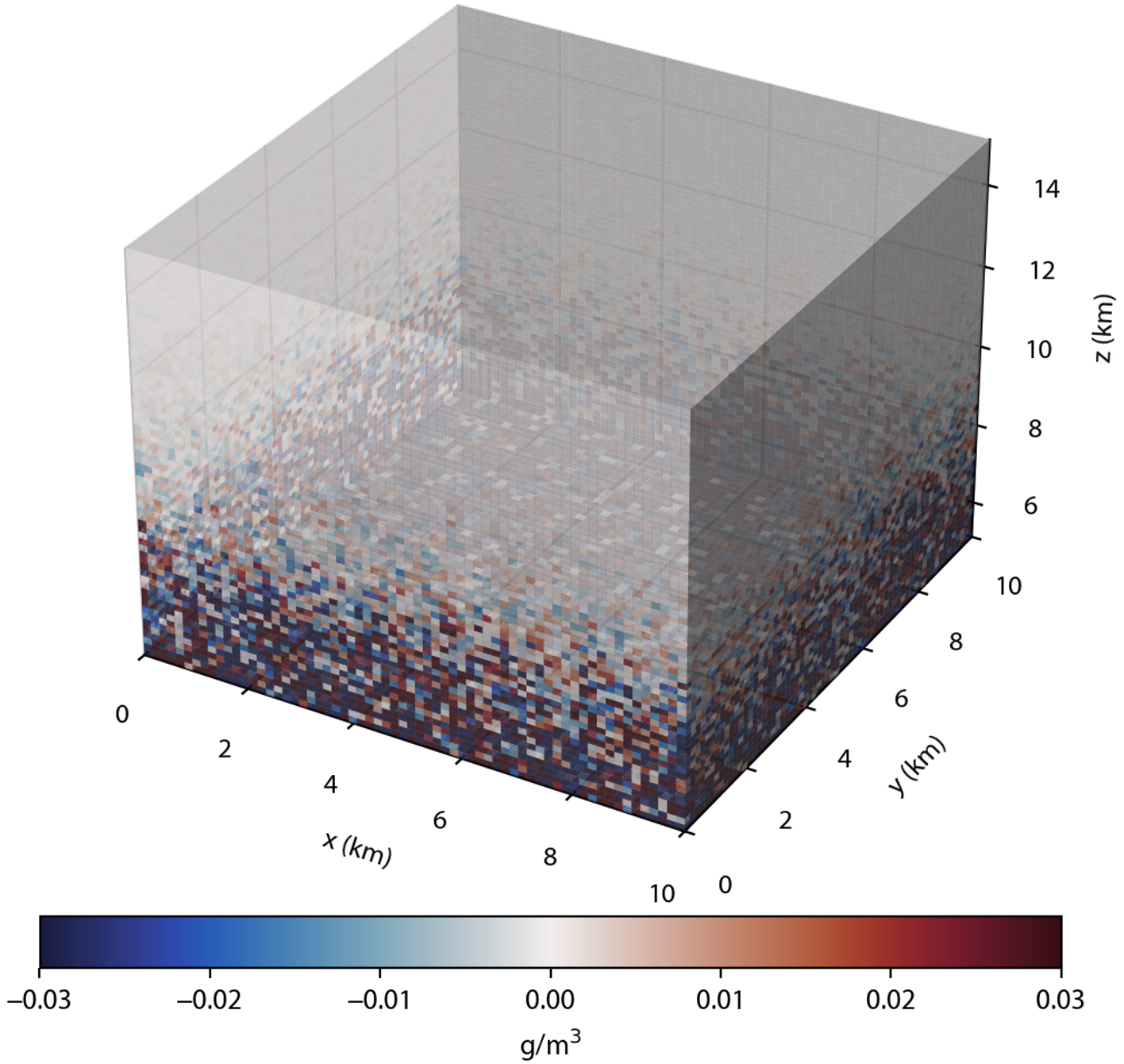}
    \captionof{figure}{WVMR fluctuations generated with AtmoCube (down-sampled to \(64 \times 64 \times 64\) cells for visualization purposes).}
    \label{fig:3}
\end{center}

%\begin{figure}[tbp]
%\centering\includegraphics[width=\textwidth]{image3.png}
%\caption{WVMR fluctuations generated with AtmoCube (down-sampled to
%\(64 \times 64 \times 64\) cells for visualization purposes).}
%\label{fig:3}
%\end{figure}

\hypertarget{wiener-filtering-of-lidar-data}{%
\subsection{Wiener-filtering of lidar
data}\label{wiener-filtering-of-lidar-data}}

The signal to noise ratio (SNR) of the water vapor fluctuations
measurement with a lidar decreases with altitude and with angular scale
along the scans. To maximize the total SNR of the lidar measurement, for
each altitude bin, we generate a time stream of PWV fluctuations as
measured by the lidar by integrating up to that bin. We then performed an
altitude-dependent Wiener filter to maximize the SNR for each Fourier
mode of these integrated time streams. Finally, for estimating each
Fourier mode of the scan, we select the estimate from the altitude bin
that maximizes the SNR of the measurement of the millimeter signal for
that mode. We recombine all Fourier modes obtained in this way to get
the final lidar estimate of the PWV fluctuation signal along the scan.

\hypertarget{result}{%
\subsection{Result}\label{result}}

The result, shown for one scan in Figure 4, is very encouraging. The
total variance of the time-domain signal is reduced by an order of
magnitude after correction using lidar data.

The simulation setup developed here can be used in the future to jointly
optimize the lidar system, the millimeter
telescope scanning strategy, and the joint data analysis pipeline for the best possible
correction as a function of astronomy science goals.

\begin{center}
    \nopagebreak % Keeps figure attached to surrounding context if needed
    \includegraphics[width=\linewidth]{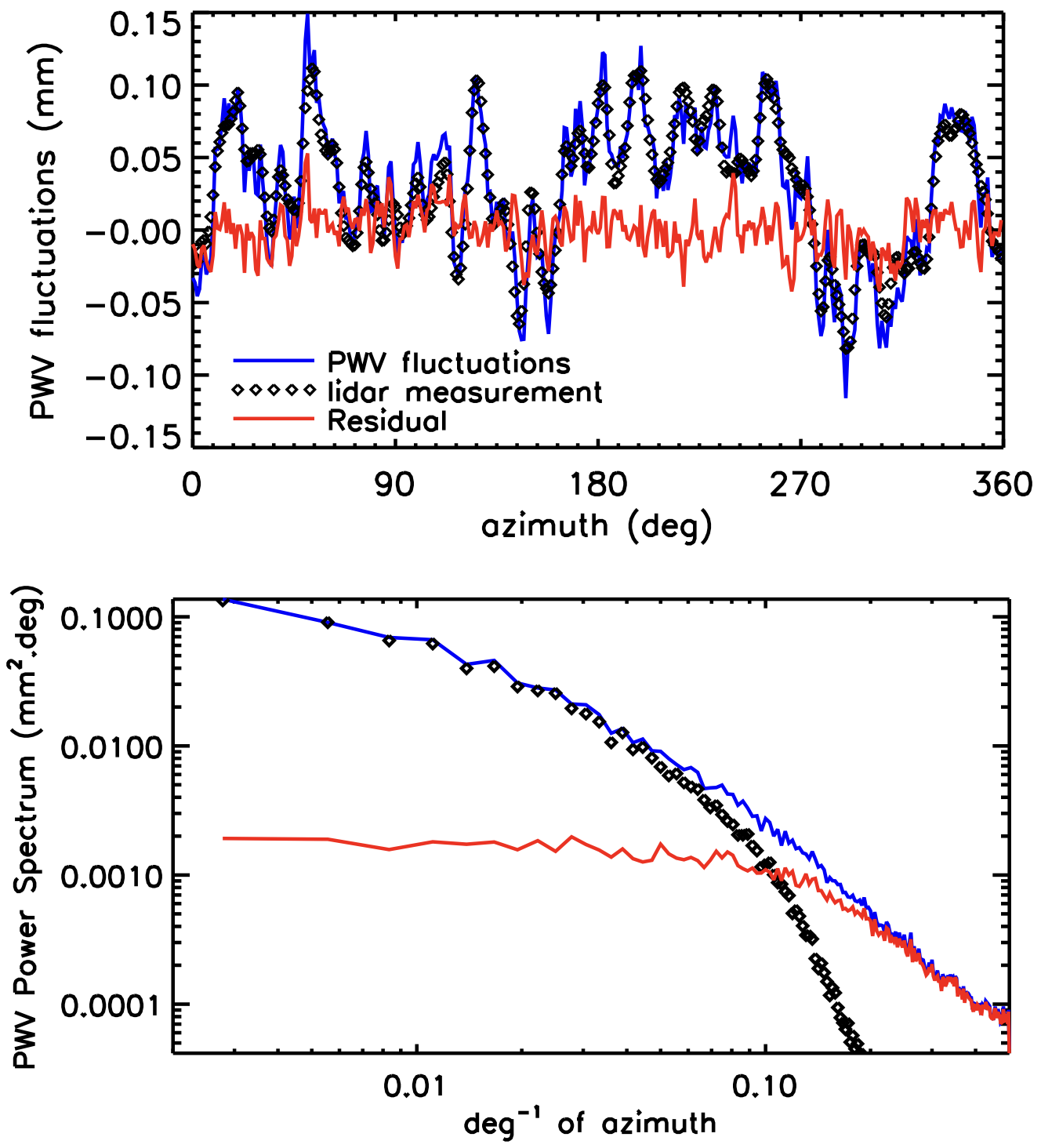}
    \captionof{figure}{Simulated measurement of scans of PWV fluctuations with a lidar. Top: data stream for one azimuthal scan; Bottom: Average power spectra computed for 100 independent scans. Blue: PWV fluctuation at mm-wavelength; black: lidar data; red: residual after correction.}
    \label{fig:4}
\end{center}
%\begin{figure}[tbp]
%\centering\includegraphics[width=0.49\textwidth]{image4.png}
%\caption{Simulated measurement of scans of PWV fluctuations with a
%lidar. Top: data stream for one azimuthal scan; Bottom: Average power
%spectra computed for 100 independent scans. Blue: PWV fluctuation at
%mm-wavelength; black: lidar data; red: residual after correction.}
%\label{fig:3}
%\end{figure}

\begin{center}
    \nopagebreak % Keeps figure attached to surrounding context if needed
    \includegraphics[width=\linewidth]{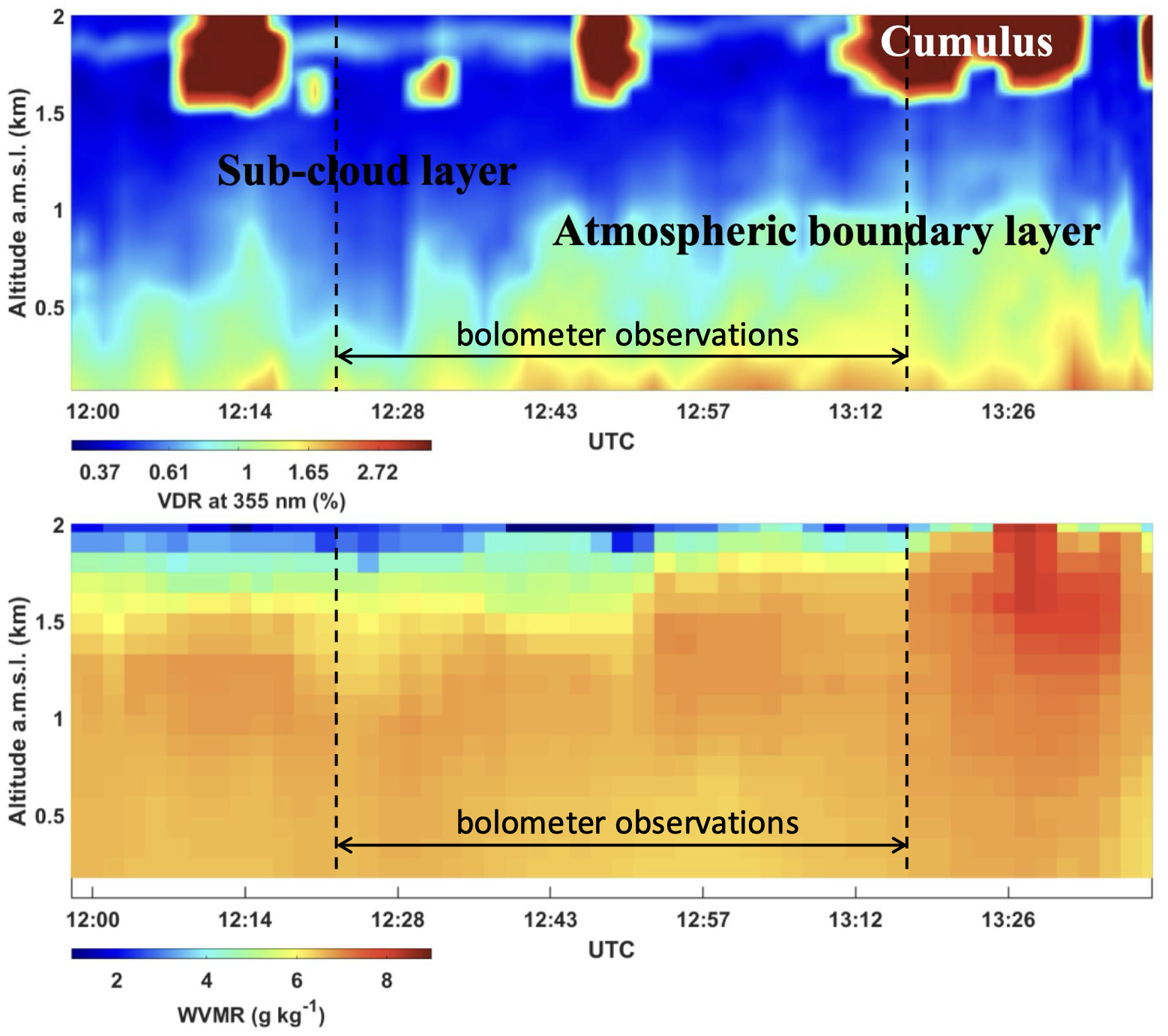}
    \captionof{figure}{Temporal evolution on 7 August 2024 of the linear volume depolarization ratio (VDR) and the WVMR.}
    \label{fig:5}
\end{center}
\hypertarget{field-campaign-in-paris}{%
\section{Field campaign in Paris}\label{field-campaign-in-paris}}

We consolidate this simulation-based proof of concept using real data.
We took advantage of the installation of the WALI lidar system {[}8{]}
pointing at zenith from the Paris Observatory in Summer 2024 (Figure 5)
to make concurrent observations with a co-pointed millimeter-wave
instrument that uses a bolometer cooled to 4~K, integrating in a broad
frequency band between \textasciitilde30 and \textasciitilde300 GHz
(shown in Figure 6).

\begin{center}
    \nopagebreak % Keeps figure attached to surrounding context if needed
    \includegraphics[width=\linewidth]{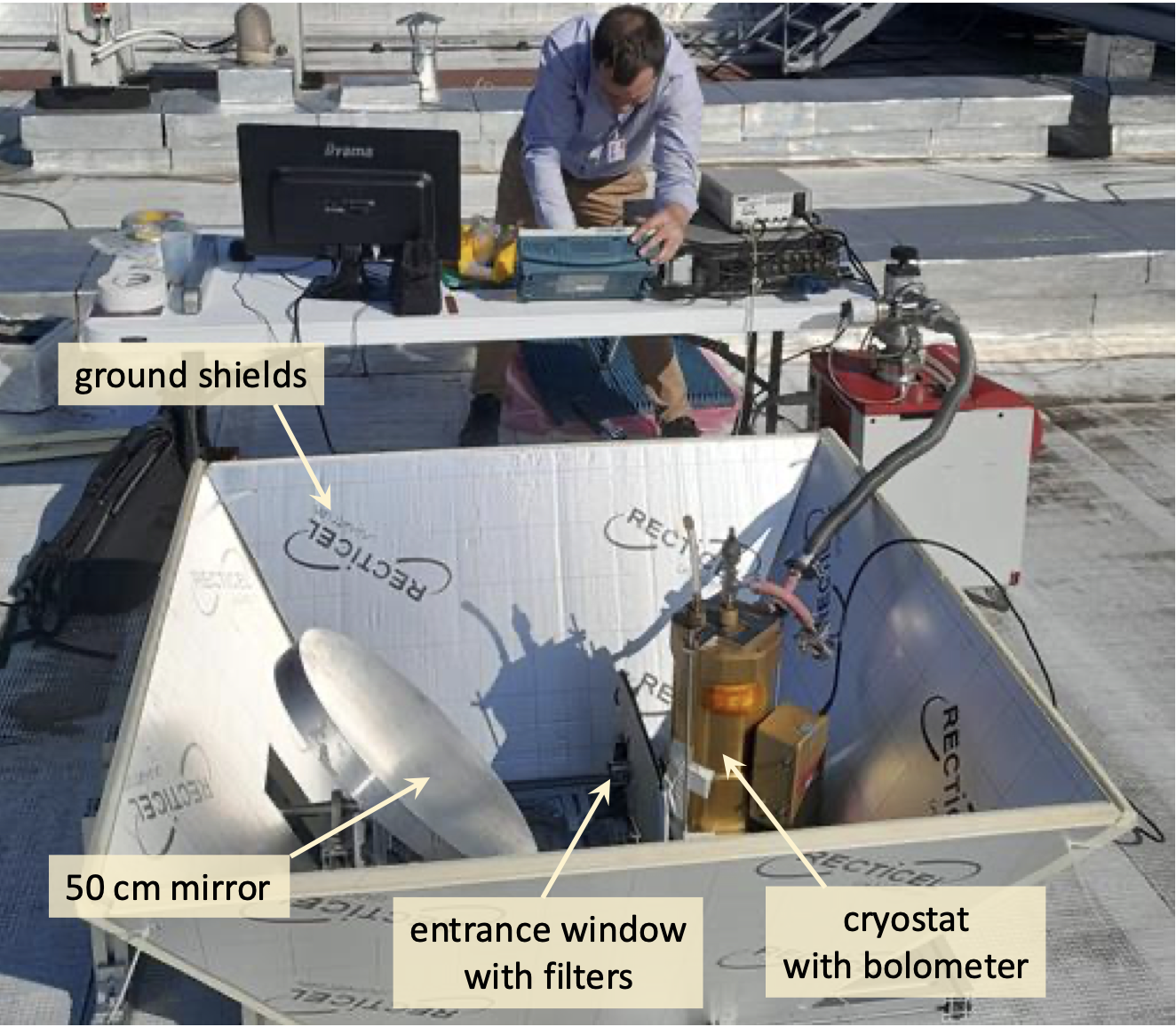}
    \captionof{figure}{Testbed millimeter telescope used for the proof-of-concept field campaign in Paris.}
    \label{fig:6}
\end{center}

%\begin{figure}[tbp]
%\centering\includegraphics[width=\textwidth]{image5.png}
%\caption{Temporal evolution on 7 August 2024 of the linear volume
%depolarization ratio (VDR) and the WVMR.}
%\label{fig:5}
%\end{figure}

%\begin{figure}[tbp]
%\centering\includegraphics[width=\textwidth]{image6.png}
%\caption{Testbed millimeter telescope used for the proof-of-concept
%field campaign in Paris.}
%\label{fig:6}
%\end{figure}

Observed drifts in the mm-wave data, fitted with a linear combination of
the lidar water vapor and VDR data streams (Figure 7), show that the
general trend in the bolometer data stream can be inferred from lidar
observations.

\hypertarget{conclusion-and-perspectives}{%
\section{Conclusion and
perspectives}\label{conclusion-and-perspectives}}

These very encouraging initial results provide a strong motivation to
further investigate how synergistic observations by atmospheric lidars
and mm-wave telescopes can be used to improve astronomical observations
and better model the turbulent processes in the Earth\textquotesingle s
atmosphere that hinder those observations. This can be a game-changer
for measurements used to trace the origins of the Universe, with
potential improvements of the sensitivity by an order of magnitude or
more for large angular scales in particular. However, this work is
still in the early stages. Atmosphere modeling must be updated
considering contributions from temperature fluctuations and ice
crystals, informed by additional lidar observations. We then plan to
consolidate our conclusions with comprehensive simulations and
comparisons with recent millimeter-wave telescope data taken under real
astronomical observing conditions.

\vspace{0.5cm}
\begin{center}
    \nopagebreak % Keeps figure attached to surrounding context if needed
    \includegraphics[width=\linewidth]{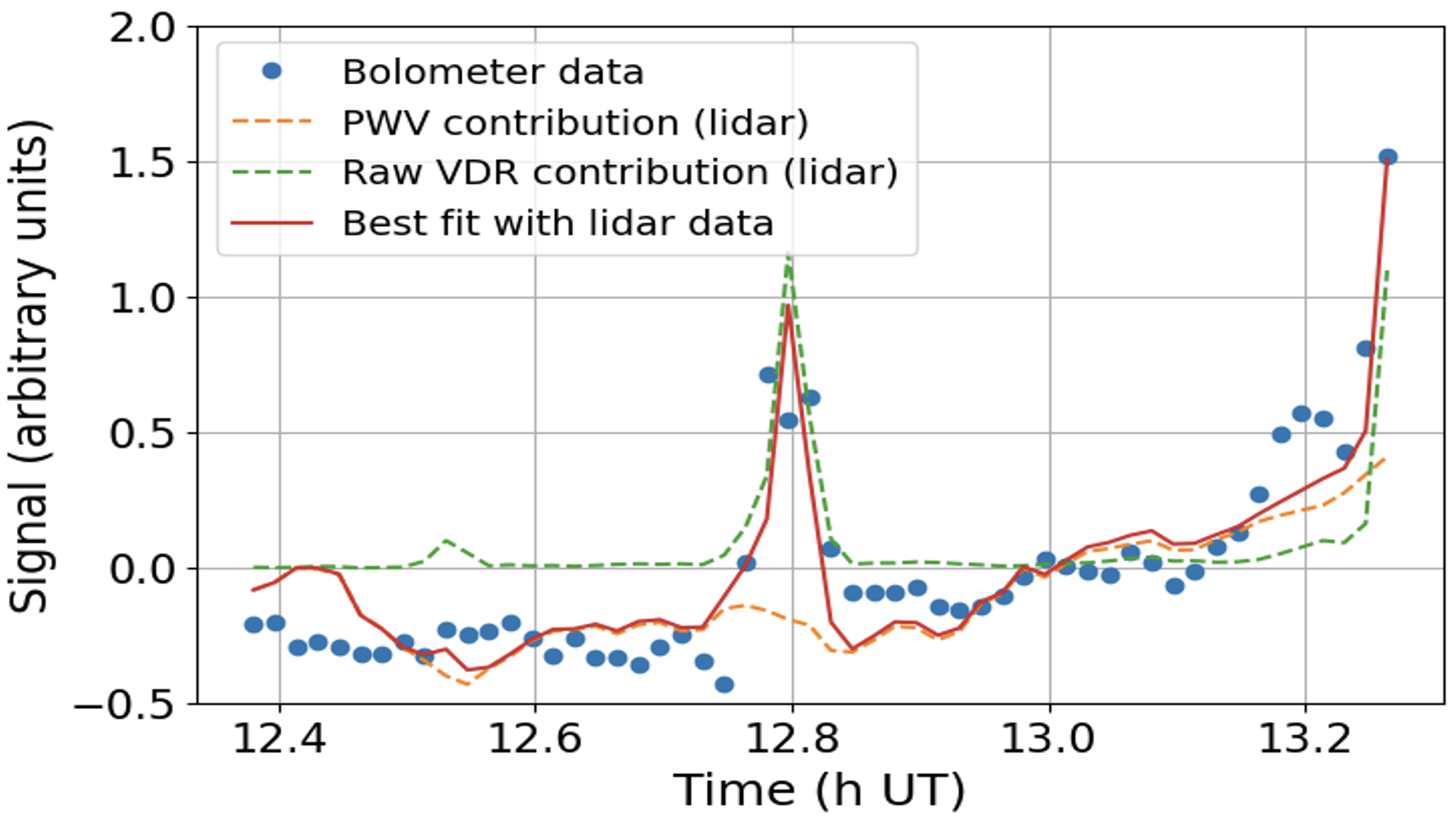}
    \captionof{figure}{A fit of bolometer data at millimeter wavelength with a linear
combination of lidar VDR and PWV data.}
    \label{fig:7}
\end{center}

%\begin{figure}[tbp]
%\centering\includegraphics[width=\textwidth]{image7.png}
%\caption{A fit of bolometer data at millimeter wavelength with a linear
%combination of lidar VDR and PWV data.}
%\label{fig:7}
%\end{figure}

\hypertarget{acknowledgements}{%
\section{Acknowledgements}\label{acknowledgements}}

We thank the Paris Observatory for hospitality and support in the summer
of 2024, and the CEA/DRF and CNRS/INSU--IN2P3 for making this study
possible.

\hypertarget{references}{%
\section{References}\label{references}}

\begin{enumerate}
\def\labelenumi{\arabic{enumi}.}
\item
  P. Ade et al. (the SO collaboration), "The Simons Observatory: Science
  goals and forecasts", JCAP, Issue 02, id. 056 (2019).
\item
  CCAT-Prime Collaboration, "Science Goals and Forecasts with Prime-Cam
  on the Fred Young Submillimeter Telescope", ApJ Supplement Series,
  Volume 264, Issue 1, id.7 (2023)
\item
  J. Errard et al., "Modeling atmospheric emission for CMB ground-based
  observations", ApJ Volume 809, Issue 1, article id. 63 (2015)
\item
  T.W. Morris et al. "The Atacama Cosmology Telescope: Quantifying
  atmospheric emission above Cerro Toco", Phys. Rev. D, Volume 111,
  Issue 8, id.082001 (2025)
\item
  P. Chazette, F. Marnas, and J. Totems, ``The mobile Water vapor
  Aerosol Raman LIdar and its implication in the framework of the HyMeX
  and ChArMEx programs: application to a dust transport process'',
  Atmos. Meas. Tech., 7, 1629--1647 (2014)
\item
  K. Abazajian et al. (CMB-S4 collaboration), "CMB-S4: Science Case,
  Reference Design, and Project Plans", arXiv:1907.004473 (2019)
\item
  H. Hersbach et al., "The ERA5 global reanalysis", Quarterly Journal of
  the Royal Meteorological Society, 146, (2020)
\item
  P. Chazette, J. Totems, ``Lidar Profiling of Aerosol Vertical
  Distribution in the Urbanized French Alpine Valley of Annecy and
  Impact of a Saharan Dust Transport Event'', Remote Sens., 15, 1070
  (2023)
\end{enumerate}
\end{multicols}

\end{document}